# Dark Matter in the Central Region of the Spiral Galaxy NGC 4321

Israa Abdulqasim Mohammed Ali[1,*], Chorng-Yuan Hwang[2,*] and Zamri Zainal Abidin[1,*]

[1]Physics Department, University of Malaya, 50603 Kuala Lumpur, Malaysia.

[2]Graduate Institute of Astronomy, National Central University, Jhongli, 32001, Taiwan.

**ABSTRACT**

We present our results of 12 CO(1-0) transition in the central region of NGC 4321 using the Atacama Large Millimeter and Sub-millimeter Array (ALMA). We found an unaccounted mass of $2.3 \times 10^9$ M$_\odot$ within the central 0.7 kpc of this galaxy. The expected mass of the supermassive black hole (SMBH) in this galaxy is much smaller than the unaccounted mass. The invisible mass is likely caused by dark matter in the central region of the galaxy, indicating a cuspy dark matter profile. We also investigated the Modified Newtonian Dynamics (MOND) as an alternative mechanism to explain the invisible mass. We noted that at the radius of 0.7 kpc of the galaxy, the acceleration is about $1.04 \times 10^{-7}$ cm s$^{-2}$, which is much larger than the critical acceleration $a_0 \sim 1.2 \times 10^{-8}$ cm s$^{-2}$ in the MOND theory, suggesting that theory might not be able explain the unseen mass problem in central region of this galaxy.

**Keywords:** cosmology; dark matter; galaxy.

*E-mail: israa.aq@siswa.um.edu.my / israa.aq88@gmail.com (IAMA);

hwangcy@astro.ncu.edu.tw (C-YH); zzaa@um.edu.my (ZZA)

# 1. INTRODUCTION

Dark matter is one of the greatest unsolved mysteries of the universe. In the standard ΛCold Dark Matter (ΛCDM) cosmology, about 25% of the mass-energy of the Universe is made up of the dark matter, which dominates gravitational evolution on large scales [1, 2]. Dark matter is also a very important component in galaxies. It is well known that the distribution of the dark matter in the galaxy defines the formation, evolution and dynamics of the galaxy. However, dark matter has little interaction with ordinary matter, and hence it is very difficult to detect [e.g., 3].

As dark matter particles have not yet been observed, the existence of dark matter remains purely theoretical. Some alternative gravity theories, such as the Modified Newtonian Dynamics (MOND)[4, 5, 6], also claim to explain the observed gravitational fields without resorting to any dark matter. MOND introduced a universal acceleration constant $a_0 \approx 1.2 \times 10^{-8}$ cm s$^{-2}$ and suggested that the dark matter phenomena in galaxies only occur in the small acceleration regime with $a \ll a_0$[7]. MOND was able to explain the rotation curves of galaxies of various luminosities [8] and was also successful in describing the dynamics of galaxy groups and clusters [9] and globular clusters [10].

A normal galaxy is thought to locate inside a much larger dark halo. The dynamics of the outer regions of the galaxy is mainly dominated by the dark matter halo. However, the dynamics of the inner regions of the galaxy is much complicated because the dynamics is also affected by the dark matter distribution, the central super massive black holes, and the stellar bulges in the central regions of the galaxy. Besides, the acceleration in the inner regions of a galaxy is expected to be rather strong because of its relatively small size and mass concentration. Therefore, the inner regions of galaxies might provide us a platform to distinguish the influence of dark matter from that of MOND.

M100 (=NGC 4321) is one of the brightest spiral galaxies (SAB(s)bc galaxy) in the Virgo cluster. M100 is tilted nearly face-on as observed from Earth with relative proximity (16.1 Mpc; Ferrarese et al. [11]). M100 has two symmetric and well-defined spiral arms. The nucleus of the galaxy is bright and compact; the galaxy was also classified as an HII/LINER [12]. M100 has received a lot of observational and theoretical attention because its moderate inclination and proximity, which make it easier to investigate the content, distribution, and kinematics of its interstellar medium and stellar components of this galaxy; the circumnuclear region of NGC 4321 has also been observed with various wavebands [e.g., 13, 14, 15].

The aim of this work is to estimate the mass distribution in the central region ($r \leq 0.7$ kpc) using the CO(1-0) observations of the Atacama Large Millimeter and Sub-millimeter Array (ALMA). The data reduction and analysis are presented in section 2. We present our results in section 3 and 4. Conclusions of our investigations are given in the last Section.

## 2. DATA SELECTION
### 2.1. ALMA Data

The ALMA data of NGC 4321 were obtained from the ALMA science verification (SV) data. The galaxy was mapped at the CO (J=1–0) transition within the Band 3 receiver of the ALMA. The resulting beam size of the CO image of NGC 4321 is $\approx 3.87'' \times 2.53''$. The Image analysis was done using the Common Astronomy Software Application (CASA) reduction package version 4.1. Fig. 1 shows the channel maps of the CO Line in the central region of NGC 4321. The channel width is 5 km s$^{-1}$, and the channel noise is about 16 mJybeam$^{-1}$. The integrated intensity (i.e., mom0) map of the CO line of NGC 4321 is shown in Fig. 2a.

### 2.2. VLA Data

The HI data of NGC 4321 were obtained from the public archive of the Very Large Array (VLA). The source was observed on 25-30 March 2003 with total integration time of 451560 seconds. The observations were carried with the D configuration of 28 antennas at the L band, and the spectra were centered at the HI 21-cm line with 63 independent channels in two polarizations (LL and RR) and with a total bandwidth of 2.71 MHz. Imaging processing was also performed with the CASA software using the task "clean" with briggs weighting. The restoring beam is $31.11'' \times 28.10''$ with PA = $-26°$. Fig. 2b shows the integrated-intensity image of the HI data, which would be used to calculate the HI mass of the galaxy ($M_{\rm HI}$).

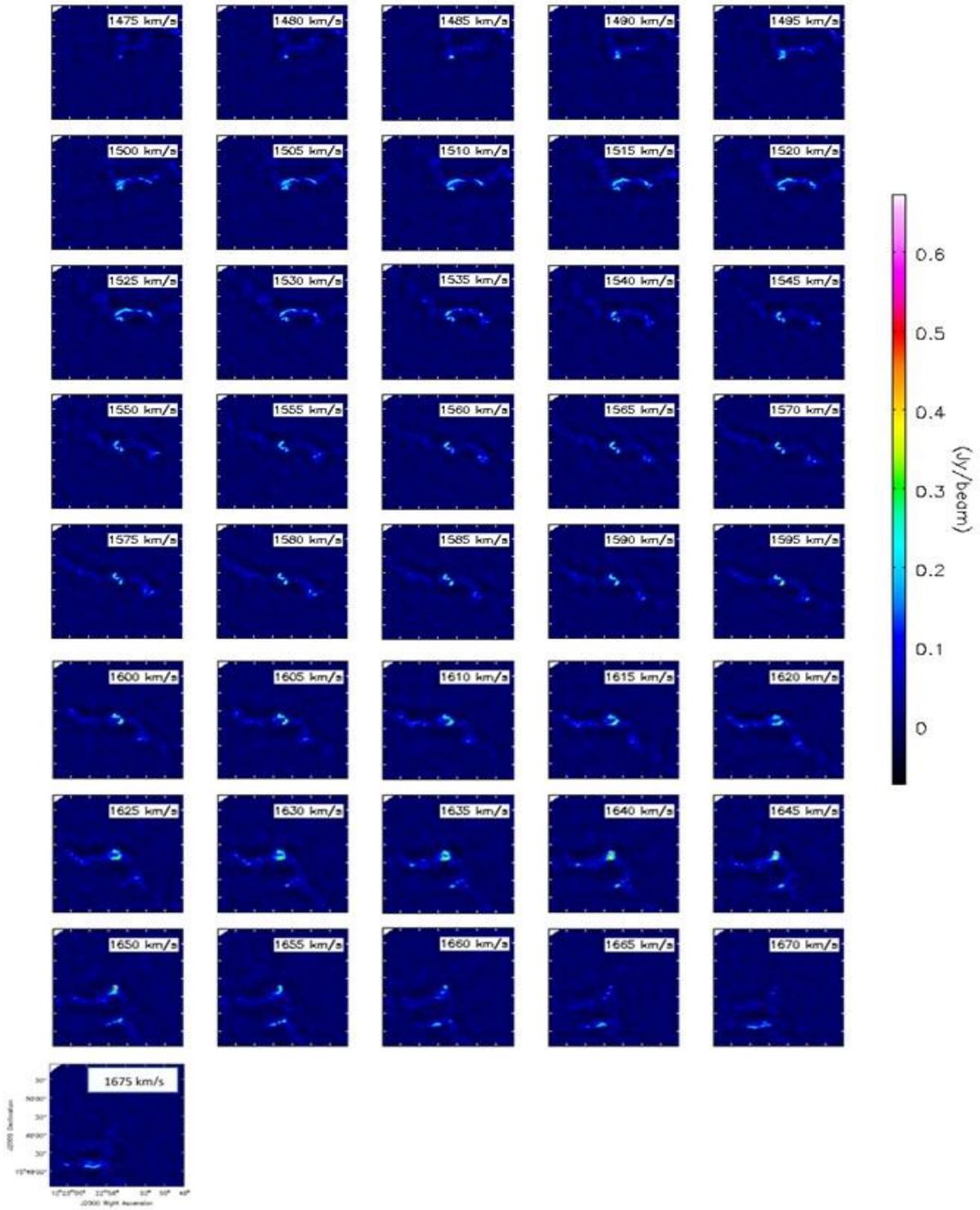

**Fig.1.** Channel maps of the CO(1-0) line emission in the central region of NGC 4321. The color scale range is shown in the wedge at right in units of Jybeam$^{-1}$.

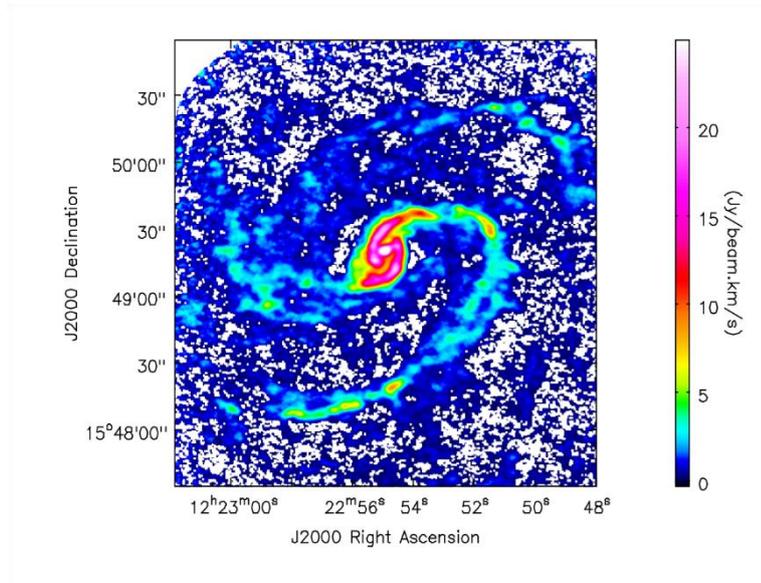

(a)

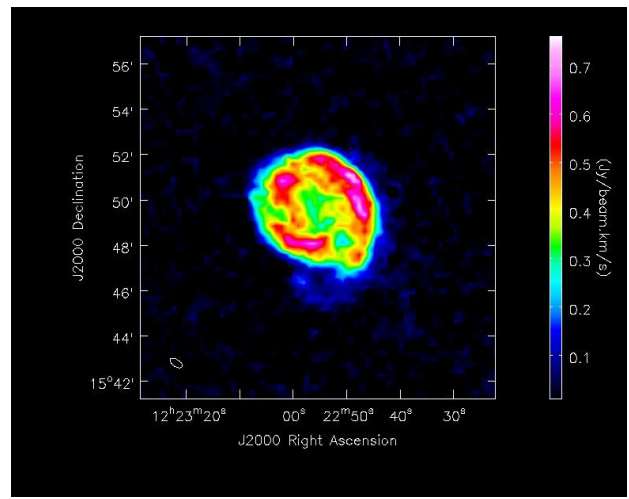

(b)

**Fig.2.** (a) Integrated intensity (mom0) map of the CO(1-0) line emission of NGC 4321. The color scale range is shown in the wedge at right in units of Jybeam$^{-1}$km s$^{-1}$. The synthesized beam is $3.87''\times2.53''$. (b) Integrated intensity of the HI observation of NGC 4321. The color scale range is shown in the wedge at right in units of Jybeam$^{-1}$km s$^{-1}$. The synthesized beam is $31.11'' \times 28.10''$.

## 3. Results

### 3.1. Rotation Curve

We derived the rotation curves from the CO data cube with the 3D code $^{3D}$Barolo [16]. This software builds tilted-ring models and compares them directly in the 3D observational space. This ensures full control of the observational effects and in particular a proper account of beam smearing that can strongly affect the derivation of the rotational velocities in the inner regions of galaxies (see [17]). The model is described by three geometrical parameters, which are the coordinates of the galaxy centre (x0, y0), the inclination i, and PA, and three kinematic parameters, which are the redshiftz, the rotation velocity 75 $V_{rot}$, and the velocity dispersion $\sigma$. To obtain rotation curve of NGC 4321, we made the following assumptions. We fixed the systemic velocity as 1575 km s$^{-1}$ from Knapen et al. [18] and all the rings were centered on the centre of NGC 4321 taken from [19] (see Table 1). Given the above assumptions, we are left with four fitting parameters: i, PA, $V_{rot}$ and $\sigma$. To estimate a good fit of the kinematics, it is necessary to start with reasonable initial guesses for i and PA. We used $^{3D}$Barolo to estimate these initial guesses by fitting the CO cube map. We then estimated the rotation and dispersion in two stages. First $^{3D}$Barolo did a fit leaving the four parameters free. Then it fixed the geometrical parameters, regularising them with a polynomial and performing a new fit of $V_{rot}$ and $\sigma$ alone. In Fig. 3 and 4 we showed the resulting rotational velocity and the comparison between the model and the data through the position-velocity diagram along the major and the minor axes. A small figure embedded in Fig. 4(top) emphasizes the well-defined solid rotation curve along the major axis. It is obvious that the galaxy tends to show a linear increase in the rotational velocity at the inner most central regions and becomes nearly constant in the outer regions. We note that the"outer" region here is still within the central 1 kpc of the galaxy.

To investigate the non-circular motions in NGC 4321 we derive the residual 95 velocity map. According to Castillo-Morales et al. [20], most of the information on non-axisymmetric components is kept in the residual velocity. To derive the residual velocity, we obtained the model velocity field by $^{3D}$Barolo as shown in Fig. 5b, then subtracting this model from the observed velocity field to obtain the residual velocity map for NGC 4321 which is shown in Fig. 5c. The residual map shows that the largest residual velocity is about 39 km s$^{-1}$ and the standard deviation of the distribution of the residual velocity is about 15 km s$^{-1}$ within the central radius of 9 arcsec, which is much smaller than the rotational velocity (~150 km s$^{-1}$) within the same radius. We emphasize that this result does not affect the derived rotational velocity of NGC 4321.

**Table 1.** Galaxy properties for NGC 4321.

| Parameter | Value |
| --- | --- |
| R.A.(2000) | 12h22m54.9s |
| Dec.(2000) | 15°49′21″ |
| Adopted distance | 16.1 Mpc |
| Inclination | 27° |
| PA | 153° |
| Systemic velocity | 1575 km s$^{-1}$ |

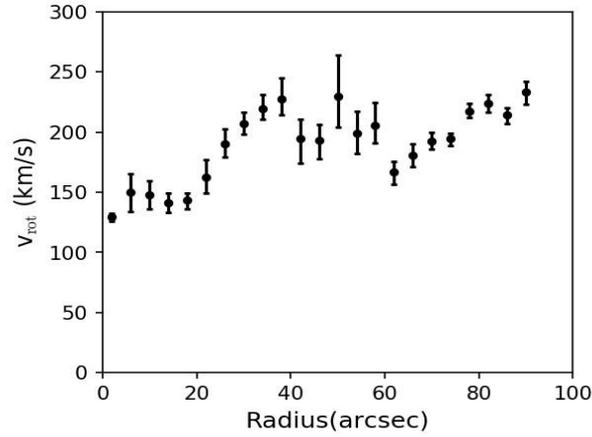

**Fig.3.** Dots with error bars are the rotation curve of NGC 4321 obtained from data cube with $^{3D}$Barolo.

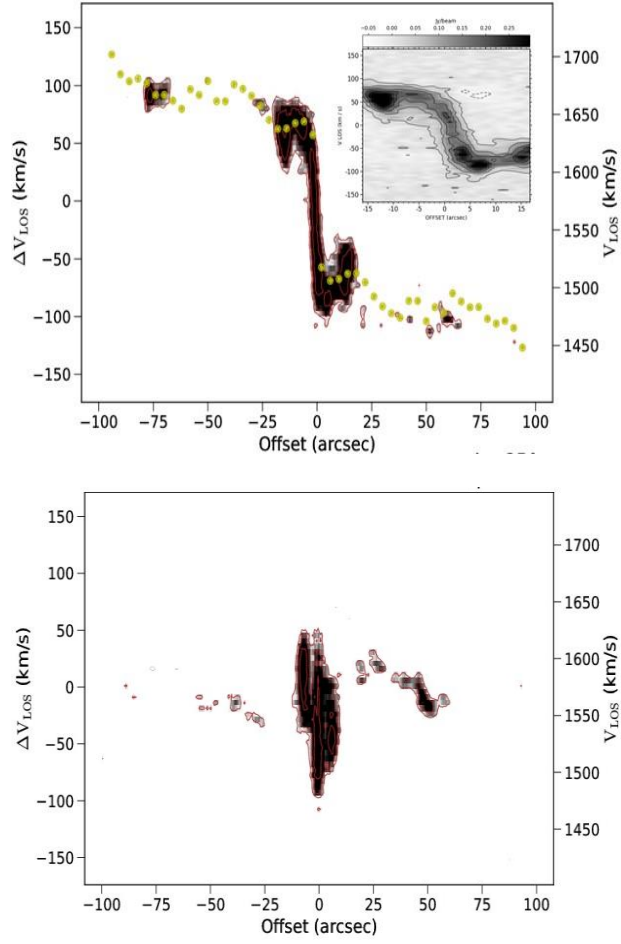

**Fig.4.** Position-velocity diagrams of NGC 4321 along major (top) and minor (bottom) axes. Data are represented in grey, model in red, rotation curve as yellow dot.

### 3.2. The Total and Gas Masses of NGC4321

The total mass of a disk galaxy can be determined from the observed rotation curve. The total mass inside a given radius R can be measured by the speed at that radius, following the Kepler formula [21].

$$M_{\mathrm{dyn}} = 2.32 \times 10^5 \times R \times V(R)^2 \ \mathrm{M_\odot} \qquad (1)$$

where $V(R)$ is the rotation velocity in km s$^{-1}$ at the radius $R$, which is in units of kpc.

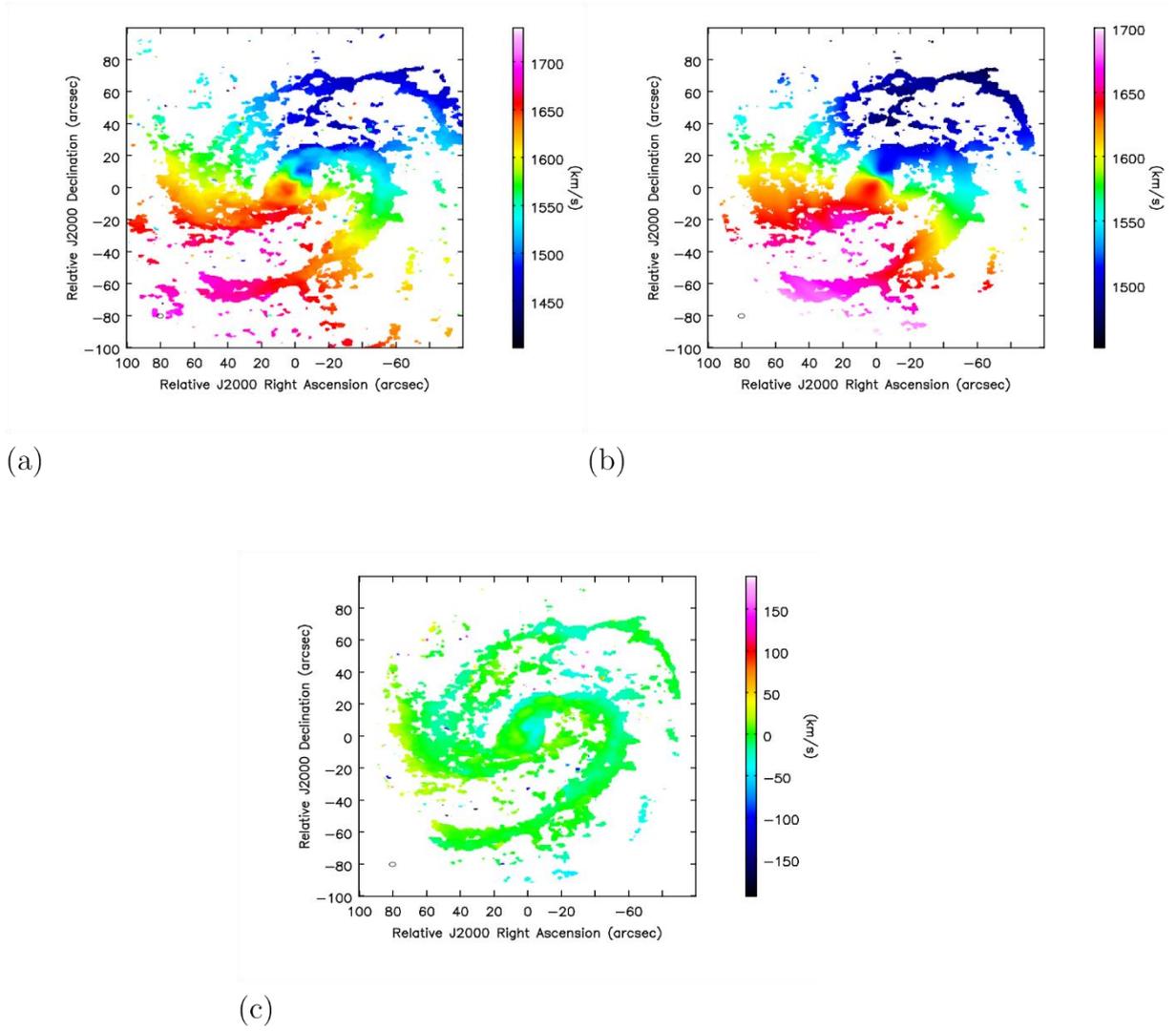

**Fig.5.** (a) velocity field of the CO(1-0) line emission of NGC 4321. (b) Model velocity field as determined from a rotation curve with $^{3D}$Barolo. (c)Residual velocity map, obtained by subtracting the model(b) from the velocity field(a). The color scale range is shown in the wedge at right in units of km s$^{-1}$. The synthesized beam is $3.87'' \times 2.53''$.

In nearly face-on galaxy (i<40) as it is the case of NGC 4321 the estimation of the V$_{rot}$ and the inclination angle are difficult because they are strongly degenerate. Moreover, for low inclination angles a small error in the estimation of the inclination implies a large uncertainly on V$_{rot}$. Since our aim is to measure the mass through the rotation curve this needs to be taken into account. We suggested the estimation of three different rotation curves, which are

(1) using the best inclination angle (i= 27°) estimated by $^{3D}$Barolo, (2) estimating an upper limit for $V_{rot}$ running Barolo with the inclination angle fixed to 20°, and (3) estimating a lower limit for $V_{rot}$ by running $^{3D}$ Barolo with inclination angle fixed to 40°. Then we can use these three different rotation curves to have three different estimation of the mass inside 9. In this way we can obtain a realistic estimate of the uncertainties. The dynamic mass $M_{tot}$ is estimated to be around $4 \pm 1.32 \times 10^9$ $M_\odot$ within the central radius of 0.7 kpc (9″).

The gas mass can be obtained from the following formula [22]:

$$M_{gas} = 1.36[M_{HI} + M_{H_2}] \; M_\odot \qquad (2)$$

where the factor 1.36 is a constant to include the contributions of He and the other heavier elements to the gas mass.

The mass of atomic neutral hydrogen was obtained from the HI 21 cm emission line (see Fig. 2). The HI mass can be derived from the HI flux using the equation [23]:

$$M_{HI} = 2.36 \times 10^5 \times D^2 \times F_{HI} \; M_\odot, \qquad (3)$$

where $D$ is the distance to the galaxy in Mpc and $F_{HI}$ is the integrated HI flux in units of Jy km s$^{-1}$. Since the resolution of the HI image is about 30″, we estimated the HI flux within the central 1′ region of the galaxy. The flux was found to be around 2.36 Jy km s$^{-1}$, which corresponds to an HI mass of $1.4 \times 10^8$ $M_\odot$. This is the total HI mass within the central 1′ region and is the upper limit of the HI mass within the central region of our investigation.

We derive the molecular gas mass, $M_{H2}$ from the observed ALMA CO flux [24]:

$$M_{H_2} = 1.2 \times 10^4 \times D^2 \times F_{CO(1-0)} \times \frac{X_{CO}}{3 \times 10^{20}}, \qquad (4)$$

where $X_{CO}$ is the CO-to-H2 conversion factor in units of cm$^{-2}$ (K km s$^{-1}$)$^{-1}$, and $D$ is distance to the source in units of Mpc. We adopt $X_{CO} = 2.2 \times 10^{20}$ cm$^{-2}$ (K km s$^{-1}$)$^{-1}$ from Sandstrom et al. [19]. The total flux is $408 \pm 6$ Jy km s$^{-1}$ within the central 18″ × 18″ region of NGC 4321; and the $M_{H2}$ is estimated to be $9 \pm 0.13 \times 10^8$ $M_\odot$. If we extended the region to include the CO flux within a 30″ (2.4 kpc) radius, we found the molecular mass becomes $M_{H_2} \approx 3 \times 10^9$ , which is close to the values obtained with single-dish observations [25].

This suggests that the missing flux of the ALMA observations is negligible. The total gas mass of NGC 4321 within the central 0.7 kpc radius, including atomic gas, molecular gas, and metal contributions, is thus about $1.4\times 10^9\ M_\odot$.

### 3.3. Stellar Mass

To estimate the stellar mass, we used the equation from Querejeta et al.[26], who used the Independent Component Analysis (ICA) method that can correctly remove the dust emission. They adopted mass-to-light ratio, M/L=0.6, which is based on Chabrier IMF of Meidt et al. [27].

$$M_{\text{stellar}} = 10^{8.35} \times \left(\frac{F_{3.6}}{\text{Jy}}\right)^{1.85} \times \left(\frac{F_{4.5}}{\text{Jy}}\right)^{-0.85} \times \left(\frac{D}{Mpc}\right)^2\ M_\odot, \qquad (5)$$

To obtain the stellar mass in the central region of the galaxy, we need to deproject the image profile $I(R)$ to obtain the luminosity density $j(r)$ distribution of the galaxy. We found that the Spitzer images of NGC 4321 can be well fitted with the modified Hubble profile [28]:

$$I(r) = \frac{I_0}{1+(\frac{R}{r_0})^2},$$

which has a simple analytic form of $j(r)$:

$$j(r) = \frac{j_0}{\left[1+(\frac{r}{r_0})^2\right]^{\frac{3}{2}}},$$

where $I_0$ is the central surface brightness, $r_0$ is the core radius, and $r_0$ and $j_0$ is related to the central surface brightness with $I_0 = 2r_0 j_0$. Fig. 6 shows the brightness profile of the Spitzer images at 3.6 $\mu$m and 4.5 $\mu$m of NGC 4321 and the fitted modified Hubble profile. The deprojected total luminosity from the central region within a radius $R$ can be derived as $L = \int_0^R 4\pi r^2 j(r) dr$. Thus, the derived stellar mass is about $3\times 10^8 M_\odot$ within the central 0.7 kpc of 160 NGC 4321.

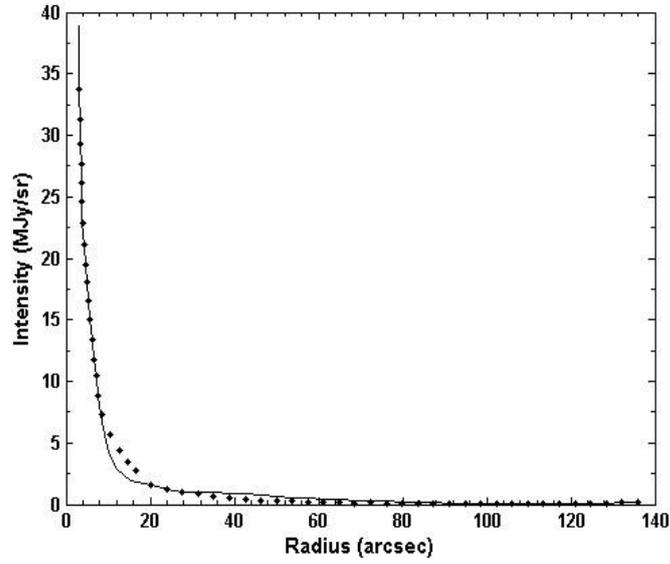

(a)

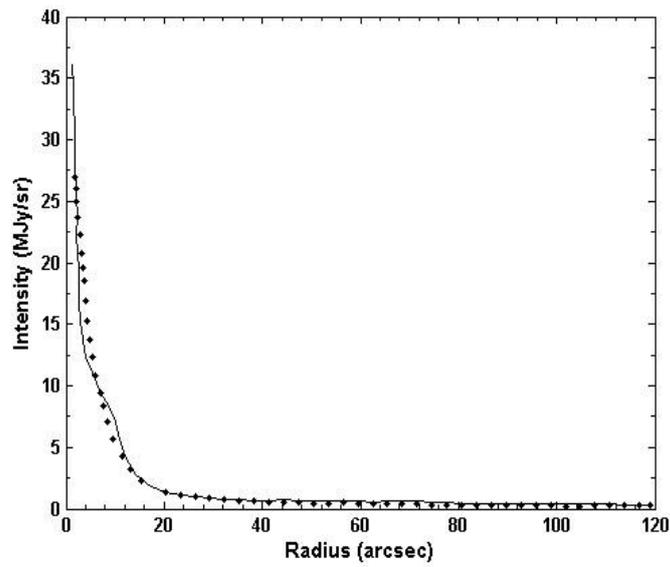

(b)

**Fig.6**. Surface brightness profile of NGC 4321 in Spitzer image. The dash line (fitting plot) is the modified Hubble law match. (a) Surface brightness profile for 3.6 $\mu$m image with $r_0 =$ 4.2 arcsec and $I_0 \approx 33.75$ MJy/sr, (b) Surface brightness profile for 4.5 $\mu$m image with $r_0 = 4.2$ arcsec and $I_0 \approx 26.94$ MJy/sr.

### 3.4. Dust Mass

Star forming galaxies might contain significant amount of dust. To estimate the dust mass in

NGC 4321, we used the IRAS 60μm and 100μm flux densities to estimate the dust mass with the following relation [29]:

$$M_{dust} = 0.959 S_{100} D^2 \left[ \left( 9.96 \frac{S_{100}}{S_{60}} \right)^{1.5} - 1 \right] M_\odot, \quad (6)$$

where $S_{60}$ and $S_{100}$ are the IRAS 60μm and 100μm flux densities in units of Jy, respectively, and $D$ is the distance to the galaxy in Mpc. The derived $M_{dust}$ of NGC 4321 is only about $2 \times 10^6\ M_\odot$. The dust mass is very small and is negligible compared with the stellar and gas mass. The dust mass of NGC 4321 was also obtained by using the integrated MIPS data at 24μm, 70μm, and 160μm using the relation given by [30]. The dust mass is similar to the result of $M_{dust}$ from the IRAS flux densities at 60μm and 100μm.

## 4. Dark Matter in the Central Region of NGC 4321

The dynamical mass of a system should include all mass components, such as stars, gas, dust, SMBHs, and dark matter.

$$M_{dyn} = M_{BH} + M_{gas} + M_{stellar} + M{dust} + M_{DM} \quad (7)$$

We can derive the dark matter mass by comparing the dynamical mass with the visible and black hole mass. The supermassive blackhole mass of NGC 4321 has been derived to be around $2.5 \pm 0.2 \times 10^7\ M_\odot$ ([31]). Comparing with all the mass components we derived, we found that the invisible mass in the central region of NGC 4321 is about $2.3 \times 10^9\ M_\odot$. Most of the invisible mass should be the dark matter, therefore, the fraction of the dark matter is about 60% of the dynamical mass with the central 0.7 kpc radius of NGC 4321.

Dark matter is the dominant mass component for most galaxies and is the fundamental ingredient in determining the properties and evolution of the galaxies. However, dark matter usually dominates in the outer regions of galaxies and is not considered as an important mass component in the central region of galaxies. The discovery of a significant amount of dark matter in the central region of NGC 4321 might also have a significant impact on the MOND theories, which were proposed to explain the "missing mass problem" without using dark matter.

A MOND theory should become the normal Newtonian dynamics at high acceleration, i.e., when the acceleration $a$ is much larger than a constant $a_0$, which was found to be $\approx 1.2 \times 10^{-8} cm\ s^{-2}$. On the other hand, at the low acceleration limit, where $a \ll a_0$, the Newtonian force $F_N$ is related to the acceleration with the form [8, 32, 33, 34, 35]:

$$\mathbf{F_N} = m\mu(\frac{a}{a_0})\mathbf{a}, \tag{8}$$

where $\mathbf{F_N}$ is the Newtonian force and $\mu(x)$ is the "interpolating function". The exact form of the interpolating function is still yet to be determined, but it should have the approximations: $\mu(x) \approx 1$ for $x \gg 1$ and $\mu(x) \approx x$ for $x \ll 1$.

Two common choices for the interpolating functions have the forms [36]:

$$\mu(\frac{a}{a_0}) = \left[1 + (\frac{a_0}{a})\right]^{-1} \tag{9}$$

and

$$\mu(\frac{a}{a_0}) = \left[1 + (\frac{a_0}{a})^2\right]^{-1/2} \tag{10}$$

An implication of the MOND theory is that we would not expect to see the dark matter phenomenon in the region of high accelerations where the dynamics should be well described by the normal Newtonian law. However, we found that the acceleration in the central region of NGC 4321 is $a = v^2/r \approx 1.04 \times 10^{-7}$ cm s$^{-2}$ at $r = 0.7$ kpc (see Fig. 3). The derived $\mu(x)$ are $\approx$ 0.89 and 0.99 for both interpolating functions respectively. Both $\mu(x)$ are too large to account for the missing mass. Therefore, it might be impossible to explain the mass discrepancy in the centre region of NGC 4321 with the traditional MOND theory. It is worth noting that there is significant amount of dark matter in the galaxy out to 0.9kpc from our observations, but we see evidence of dark matter even in the central region and this can not be explained with MOND.

## 5. SUMMARY

We reported the mass distributions of NGC 4321 using ALMA CO(1-0) data and NIR data of Spitzer. We showed that within the central 0.7 kpc radius of the galaxy, there is a significant amount of invisible mass, about 60% of the dynamical mass, which can not be explained with the gas mass and the stellar mass within this region. We suggested that there might be a significant amount of dark matter in the central region of the galaxy. What is more important is that this missing mass problem can not be explained with a traditional MOND theory because of strong acceleration in the central region of the galaxy. Since most of the dark matter phenomena at galactic scales can usually be explained by the MOND theory very well, this discovery might thus pose a significant challenge to the traditional MOND models.

## 6. ACKNOWLEDGMENTS

This paper makes use of the following ALMA data:ADS/JAO.ALMAA# 2011.0.00004.SV. ALMA is a partnership of ESO (representing its member states), NSF (USA) and NINS (Japan), together with NRC (Canada), NSC and ASIAA (Taiwan), and KASI (Republic of Korea), in cooperation with the Republic of Chile. The Joint ALMA Observatory is operated by ESO, AUI/NRAO and NAOJ. C.-Y. H. acknowledges support from the MOST grant of Taiwan MOST 103-2119-M008-017-MY3.